\documentclass[runningheads]{llncs}
\usepackage{color,soul}
\usepackage[para]{footmisc}
\usepackage{booktabs}
\usepackage{multirow}
\usepackage{graphicx}
\usepackage{amsmath,amssymb,amsfonts}
\usepackage{hyperref}

\begin{document}

\title{$Q$-space Conditioned Translation Networks for Directional Synthesis of Diffusion Weighted Images from Multi-modal Structural MRI}

\makeatletter
\newcommand{\printfnsymbol}[1]{%
  \textsuperscript{\@fnsymbol{#1}}%
}
\makeatother

\author{Mengwei Ren\index{Ren, Mengwei}\printfnsymbol{1} \and
Heejong Kim\index{Kim, Heejong}\thanks{These authors contributed equally.} \and
Neel Dey\index{Dey, Neel}\and
Guido Gerig\index{Gerig, Guido}}

\institute{Department of Computer Science and Engineering, New York University, NY, USA }

\titlerunning{Multi-modal $q$-space Conditioned Translation Networks}
\authorrunning{M. Ren and H. Kim et al.}

\maketitle              
\begin{abstract}
Current deep learning approaches for diffusion MRI modeling circumvent the need for densely-sampled diffusion-weighted images (DWIs) by directly predicting microstructural indices from sparsely-sampled DWIs. However, they implicitly make unrealistic assumptions of static $q$-space sampling during training and reconstruction. Further, such approaches can restrict downstream usage of variably sampled DWIs for usages including the estimation of microstructural indices or tractography. We propose a generative adversarial translation framework for high-quality DWI synthesis with arbitrary $q$-space sampling given commonly acquired structural images (e.g., B0, T1, T2). Our translation network linearly modulates its internal representations conditioned on continuous $q$-space information, thus removing the need for fixed sampling schemes. Moreover, this approach enables downstream estimation of high-quality microstructural maps from arbitrarily subsampled DWIs, which may be particularly important in cases with sparsely sampled DWIs. Across several recent methodologies, the proposed approach yields improved DWI synthesis accuracy and fidelity with enhanced downstream utility as quantified by the accuracy of scalar microstructure indices estimated from the synthesized images. Code is available at \url{https://github.com/mengweiren/q-space-conditioned-dwi-synthesis}. 

\end{abstract}

\section{Introduction} 
Diffusion MRI is fundamental to \textit{in vivo} tissue microstructure characterization. Recent dMRI models such as neurite orientation dispersion and density imaging (NODDI)~\cite{ZHANG20121000} and diffusion kurtosis imaging (DKI)~\cite{jensen2005diffusional} give deeper insight into tissue configurations than traditional indices such as Fractional Anisotropy. However, advanced dMRI models require dense $q$-space sampling and prolonged acquisitions, leading to increases in motion corruption and eddy-current and susceptibility artefacts and necessitate outlier DWI exclusion and/or correction. 

Several deep network approaches propose to predict high-quality microstructure models such as NODDI/DKI from undersampled DWIs given a training set of fully sampled images~\cite{chen2019xq,chen2020estimating,gibbons2019simultaneous,qspacedl}. ~\cite{gibbons2019simultaneous,qspacedl} use deep networks to estimate DKI and NODDI parameters from only 12 and 8 DWIs, respectively. ~\cite{chen2020estimating} applies graph convolutional networks for microstructure estimation from 36 DWIs. ~\cite{lin2019fast} uses fiber orientation distribution function coefficients from fully sampled images to obtain high-quality tractography from subsampled DWIs. Image translation methods~\cite{pix2pix2017,DBLP:journals/corr/abs-1812-04948,Zhu_2017_ICCV} have also been applied towards prediction of diffusion indices from other commonly co-acquired modalities, including structural/functional MRI to DTI translation~\cite{anctil2020manifold,son2019synthesizing,tian2020deepdti} and translation to DTI-derived scalars~\cite{alexander2017image,gu2019generating}. While promising, current deep network approaches require predefined and \textit{static} $q$-space sub-sampling schemes. Practical DWI processing pipelines \cite{glasser2013minimal,nir2013effectiveness} create unpredictable sampling patterns during motion \& eddy-current correction which involves the reorientation of gradients and/or their complete exclusion. As these corruptions cannot be anticipated, DWI gradients can be degraded unpredictably and require methods for arbitrary directional restoration.

\begin{figure}[t]
    \centering
    \includegraphics[width=0.95\textwidth]{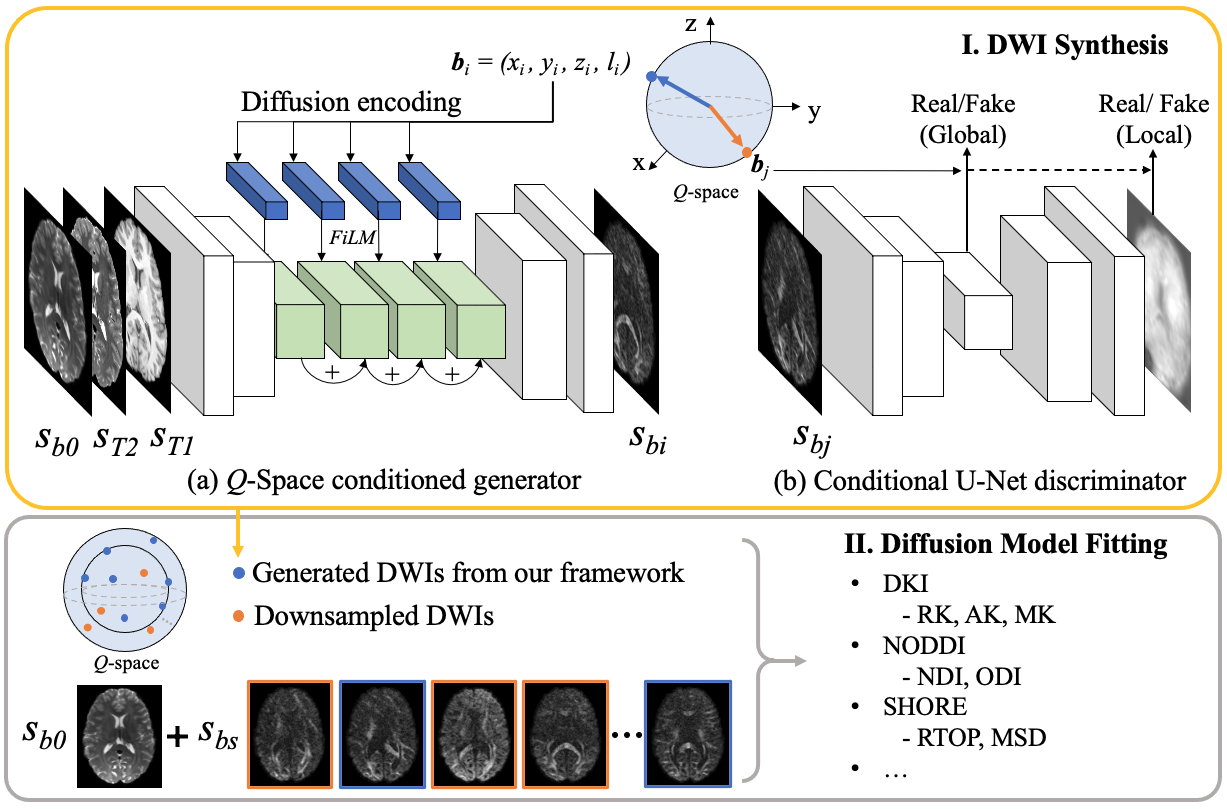}
    \caption{\textbf{Framework Overview.} \textbf{(Top)} Our $q$-space conditioned translation framework. \textbf{(Bottom)} Once trained, the generator is able to synthesize DWIs along gradients in $q$-space (blue dots). Merged with arbitrarily downsampled DWIs, various microstructural indices can be calculated with diffusion model fitting.}
    \label{fig:overview}
\end{figure}

We propose a DWI synthesis framework compatible with arbitrary $q$-space sampling by constructing a generative adversarial translation network to learn mappings between commonly co-acquired structural MRI (e.g., B0, T1, T2) to DWI with user-specified $b$-values and $b$-vectors. We incorporate diffusion sampling schemes into the translation by linearly modulating internal network representations with $q$-space coordinate information.
The synthesized DWIs can be merged with sparsely-sampled DWIs to simulate dense $q$-space acquisition to enable the usage of diffusion models which require high angular resolution. Our contributions include: (1) An adversarial learning framework for structural-to-diffusion MRI translation \textit{independent} of any dMRI model assumption and $q$-space sampling scheme; (2) $Q$-space conditioned generator and discriminator architectures which encode diffusion priors into the spatial and spectral translation framework leading to improved visual and structural fidelity over current methodologies; (3) Flexible downstream utility enabling the usage of any microstructural index, tractography strategy, and dODF estimation method. 

\section{Methods}
Fig.~\ref{fig:overview} gives an overview of our framework (yellow outline) alongside its potential downstream utility towards model fitting in downsampled $q$-space (blue outline). Once trained, the generator can generate DWIs along missing/excluded gradients in $q$-space which can be merged with the original sparsely-sampled data. \\

\noindent\textbf{Formulation.} 
The model inputs are structural images $s = (s_{b_0}, s_{T_2},s_{T_1})$, where the channels represent a baseline ($B0$) image denoted by $s_{b_0}$, and structural T2 and T1-weighted images, denoted by $s_{T_2}$ and $s_{T_1}$, respectively. Conditioned on $q$-space coordinates $\mathbf{b}=(\mathbf{\theta}, l)$ where $\mathbf{\theta} =(x, y, z)$ is a unit b-vector, and $l$ is the b-value for each shell, the generator $G: \{s_{b_0}, s_{T_2}, s_{T_1}, \mathbf{b}\} \rightarrow s_{b}$ learns to conditionally translate $s$ to a DWI $s_{b}$ supervised by a groundtruth DWI alongside a $q$-space conditioned discriminator $D$ which is trained to distinguish between synthesized and real DWI at given $q$-space coordinates. \\

\noindent\textbf{Q-space Conditioned Generator.}
As we aim to generate $s_b$ with arbitrary $\mathbf{b}$, our generator network needs to account for the continuous variable $b$ in its generation.
We do this via Feature-wise Linear Modulation~\cite{DBLP:journals/corr/abs-1709-07871} which has been shown to outperform other network conditioning mechanisms in a wide range of medical imaging tasks~\cite{jacenkow2020inside,ren2021segrenorm}. Taking structural images and desired $q$-space coordinates as input, the generator injects $q$-space information into pre-activation instance-normalized feature maps via a channel-wise linear transformation, whose parameters are learned by a 3-layer MLP from the specified $q$-space coordinates. The transformation is defined as $\gamma_{c}^{k}(\textbf{b}) \Big( \frac{h_{ c}^{k}-\mu_{c}^{k}}{\sigma_{c}^{k}} \Big) + \beta_{c}^{k}(\textbf{b})$, where $h_{c}^k$ is the feature map from the $c$th channel of $k$th convolutional layer before activation, $\mu_{c}^{k}$ and $\sigma_{n,c}^{k}$ are the mean and standard deviation of channel $c$, and $\gamma_{c}^{k}$ and $\beta_{c}^{k}$ are the scale and shift parameters conditionally learned with fully connected layers from $\textbf{b}$. Given both structural information and gradient configurations, the generator decodes and upsamples the deep features to a translated dMRI. \\

\noindent\textbf{Conditional U-Net Discriminator.}
Existing translation GANs typically use unconditional PatchGAN discriminators~\cite{pix2pix2017,Zhu_2017_ICCV}, providing only patch level feedback to the generator. We make two important changes: (1) to provide assessments of both local and \textit{global} image realism for improved synthesis, we use a U-Net discriminator~\cite{schonfeld2020u} $D$ whose bottleneck discerns global realism and whose output feature determines pixel-level realism; (2) Following recent work in conditional synthesis~\cite{DBLP:journals/corr/abs-1802-05637},
our discriminator assesses both whether the synthesized image was sampled from the real distribution and whether the synthesized image corresponds to the desired conditioning based on the $q$-space diffusion vector $\mathbf{b}$.
$D$ takes $x=(s_{b0}, s_{T_2}, s_{T_1}, s_{b_i})$ as input, where $s_{b_i}$ is either sampled from a real dMRI or a generated image $G(s_{b0}, s_{T_2}, s_{T_1}, {b_i})$, and extracts a global representation $\phi(x)$ via the encoding path $D_{enc}$ to assess global realism. Additionally, a decoder $D_{dec}$ expands $\phi(x)$ to the input size and outputs per-pixel realism feedback. To incorporate $q$-space coordinates into the discriminator, a conditional projection via inner product is inserted before the last layer of both global and local branches following~\cite{DBLP:journals/corr/abs-1802-05637}. The final layer of $D$ is defined as
$
f(x, \mathbf{b}):=\mathbf{b}^{\mathrm{T}} V \boldsymbol{\phi}\left(\boldsymbol{x}\right)+\psi\left(\boldsymbol{\phi}\left(\boldsymbol{x}\right)\right),
$
where $V$ is a learnable embedding of condition $\mathbf{b}$; $\phi(x)$ is the output before conditioning, and $\psi(\cdot)$ is a scalar function of $\phi(x)$. \\

\noindent\textbf{$Q$-space Data Augmentation}. For improved generalization, we develop two forms of DWI-physics informed augmentation: (1) With input $l = 0 s/mm^2$, the model should produce the baseline image ($B0$). We zero-out the b-value input with probability $0.1$ and replace the target image with the $B0$ image; (2) Diffusion signals from spherically antipodal directions are ideally identical. Therefore, we replace $\theta$ with $-\theta$ with probability $0.1$ and train the generator and discriminator to produce the same output. \\

\noindent\textbf{Learning Objectives.}
We use the least squares adversarial objective~\cite{DBLP:journals/corr/MaoLXLW16} for $D$ computed from both a local and global perspective as $\mathcal{L}_{GAN}(D) =  \mathcal{L}_{GAN}({D_{enc}}) +  \mathcal{L}_{GAN}({D_{dec}})$, where $\mathcal{L}_{GAN}({D_{enc}})$ and $\mathcal{L}_{GAN}({D_{dec}})$ are defined as,
\begin{align*}
\begin{split}
    &\mathcal{L}_{GAN}({D_{enc}}) = \frac{1}{2}\mathbb{E}_{y}\left[||1-D_{enc}(y, \textbf{b}_j)||^2_2\right] 
    + \frac{1}{2}\mathbb{E}_{x}\left[||D_{enc}(x, \textbf{b}_i)||^2_2\right],\\
    &\mathcal{L}_{GAN}({D_{dec}}) = \frac{1}{2}\mathbb{E}_{y}\left[\sum_{u,v}||1-D_{dec}(y, \textbf{b}_j)_{u,v} ||^2_2\right] + \frac{1}{2}\mathbb{E}_{x}\left[\sum_{u,v}||D_{dec}(x,\textbf{b}_i)_{u,v}||^2_2\right],
    \end{split},
\end{align*}
where $x=(s_i, G(s_i, b_i)), s_i=(s_{b_{0i}}, s_{T_{2i}}, s_{T_{1i}})$ is the concatenated synthesized tuple of structural inputs and generated dMRI, $y=(s_j, s_{b_j})$ is the real tuple with $s_{b_j}$ sampled from the training data and $(u,v)$ is the image-plane indices of each pixel location. 
The least-squares generator loss is defined as follows:
\begin{align*}
\begin{split}
    \mathcal{L}_{GAN}(G) = \frac{1}{2}\mathbb{E}_{x}\left[||D_{enc}(x, \textbf{b}_i) -1 ||_2^2\right] + \frac{1}{2}\mathbb{E}_{x}\left[\sum_{u,v}||D_{dec}(x,\textbf{b}_i)_{u,v} -1 ||_2^2\right],
    \end{split}
\end{align*}
To better match low-frequency details and maintain consistency with the input, we further use an intensity-based translation term as below, 
\begin{align*}
\mathcal{L}_{L_1}(G) = \left \{\begin{array}{ll}
    \mathbb{E}_{s, b}\left[\|s_{b} - G(s, \textbf{b})\|_1\right], &  \text{if } l > 0\\
    \mathbb{E}_{s, b}\left[\|s_{b_0} - G(s, \textbf{b})\|_1\right], & \text{if } l=0
\end{array}\right.
\end{align*}
where $s=(s_{b0}, s_{T2}, s_{T1})$ is the two-channel structural input, $s_b$ is the reference DWI with diffusion $\textbf{b}=(\mathbf{\theta}, l)$. 
Our complete objective function is summarized as
$\mathcal{L}(G, D) = \lambda_{GAN}\mathcal{L}_{GAN}(G,D)+ \lambda_{L_1}\mathcal{L}_1(G)$, where $\lambda_{GAN}$ and $\lambda_{L_1}$ represent the weights applied to adversarial terms and translation terms, respectively.

\section{Experiments}
\noindent\textbf{Data and preprocessing.}   
\noindent{We} experiment on the HCP 500 release~\cite{van2013wu} including preprocessed images from 19 unrelated subjects. The DWIs are obtained at 3T with b-values of $1000, 2000,$ and $3000 s/mm^2$ covering 270 gradient directions with a voxel size of $1.25 \times 1.25 \times 1.25 mm^3$. 
The prealigned T1 and T2 images are resampled to match the image resolution of B0 images. We exclude the dataset-provided corrected DWI gradients~\cite{andersson2015non} such that the gradients used for training are raw acquisitions. Consequently, each subject may have a different number of DWIs (ranging from 253 to 269). We use 9 subjects for training; 1 subject for validation and model selection; and 9 held-out subjects for final testing. \\

\noindent\textbf{Implementation details.} 
\label{sec:implementation}
The generator and discriminator learning rates are $10^{-4}$ and $5\times10^{-5}$, respectively, using Adam~\cite{kingma2014adam} optimization ($\beta_1=0.5$, $\beta_2=0.999$) and batch size $12$. Loss function weights are set to $\lambda_{GAN}$ = 1 and $\lambda_{L1} = 100$. During training, the discriminator is updated once for every two generator updates.
We train on 2D axial slices sampled from random gradient directions. DWI intensities are rescaled voxel-wise by their corresponding $B0$ image and b-values are normalized by their maximum value. All models are implemented in PyTorch 1.7.1 and trained on an Nvidia P100 GPU (12 GB vRAM). Inference time is 0.01 s/slice. Architectural implementation details are available in the supplementary material. \\

\begin{figure}[t]
    \centering
    \includegraphics[width=1\textwidth]{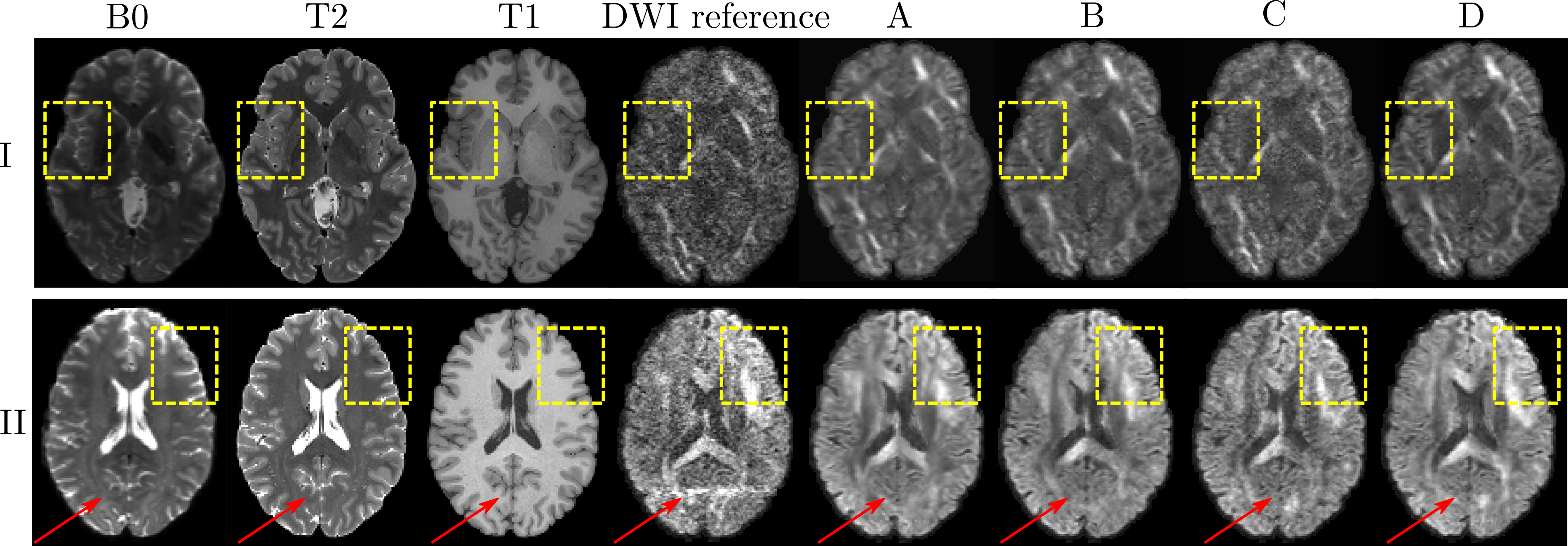}
    \caption{DWI synthesis results under various model configurations detailed in Table~\ref{tab:img-mae-ssim-psnr}.
    All potential input channels (B0, T1, T2) are visualized on the left. I. shows translation results with a standard DWI slice in the test set as the reference; II. visualizes a test slice where artefacts (red arrows) appear in the reference DWI, which are corrected by the generated DWIs from our methods. Readers are encouraged to zoom in.}
    \label{fig:dwis-vis}
    \begin{tabular}[t]{c|l|l|c|c|c}
    \hline
    Model & Input & Loss       & PSNR ($\uparrow$)  &SSIM ($\uparrow$)   &MAE ($\downarrow$)   \\ \hline
    A & b0 & L1        & $29.58$($\pm7.31e$-1) & $0.939$($\pm1.12e$-2) & $0.0563$($\pm 2.62e$-3) \\
    B & b0 & L1+GAN     & $29.65$($\pm7.61e$-1) & $0.939$($\pm1.11e$-2) & $0.0559$($\pm 2.80e$-3) \\
    C & b0+T2 & L1+GAN   & $29.64$($\pm7.38e$-1) & $0.942$($\pm1.04e$-2) & $0.0558$($\pm 2.64e$-3) \\
    D & b0+T2+T1 &L1+GAN & $\mathbf{29.77}$($\pm\mathbf{7.18e}$-\textbf{1}) & $\mathbf{0.944}$($\pm\mathbf{1.02e}$-\textbf{2}) & $\mathbf{0.0550}$($\pm \mathbf{2.28e}$-\textbf{3}) \\
    \hline
    \end{tabular}
    \caption{Quantitative DWI synthesis quality on held-out test subjects as measured by PSNR (higher is better), SSIM (higher is better), and MAE (lower is better).}
    \label{tab:img-mae-ssim-psnr}
\end{figure}

\noindent\textbf{Diffusion-weighted image synthesis.}
We compare four models (configurations A, B, C, and D in Table~\ref{tab:img-mae-ssim-psnr}) under varying loss functions and input modalities to gauge the benefits of the adversarial loss and multimodal inputs in both isolation and combination, with a summary of test set metrics calculated in 3D and averaged across subjects given in Table~\ref{tab:img-mae-ssim-psnr}. Fig. \ref{fig:dwis-vis} visualizes two example images generated from the four evaluated settings.
To assess straightforward translations of $B0$ images to DWI, Model A is constructed as a single-modality non-adversarial network. As shown in Fig.~\ref{fig:dwis-vis}A, Model A can capture overall DWI structure in its translation, but loses significant high-frequency details associated with DWI due to suboptimal objectives~\cite{pix2pix2017,larsen2016autoencoding,nilsson2020understanding}. 
On top of Model A, we add an adversarial objective for Models B, C, and D with all settings leading to improved visual fidelity as compared to Model A. Compared to Model B (which uses only $B0$ images), Models C and D show progressively enhanced structure by concatenating additional T1 and T2 channels (see yellow insets in Fig.~\ref{fig:dwis-vis}) alongside improved PSNR, SSIM and MAE (Table~\ref{tab:img-mae-ssim-psnr}). Following this investigation of model configurations, we use Models C and D for further experiments. \\

\begin{figure}[t]
    \centering
    \includegraphics[width=1\textwidth]{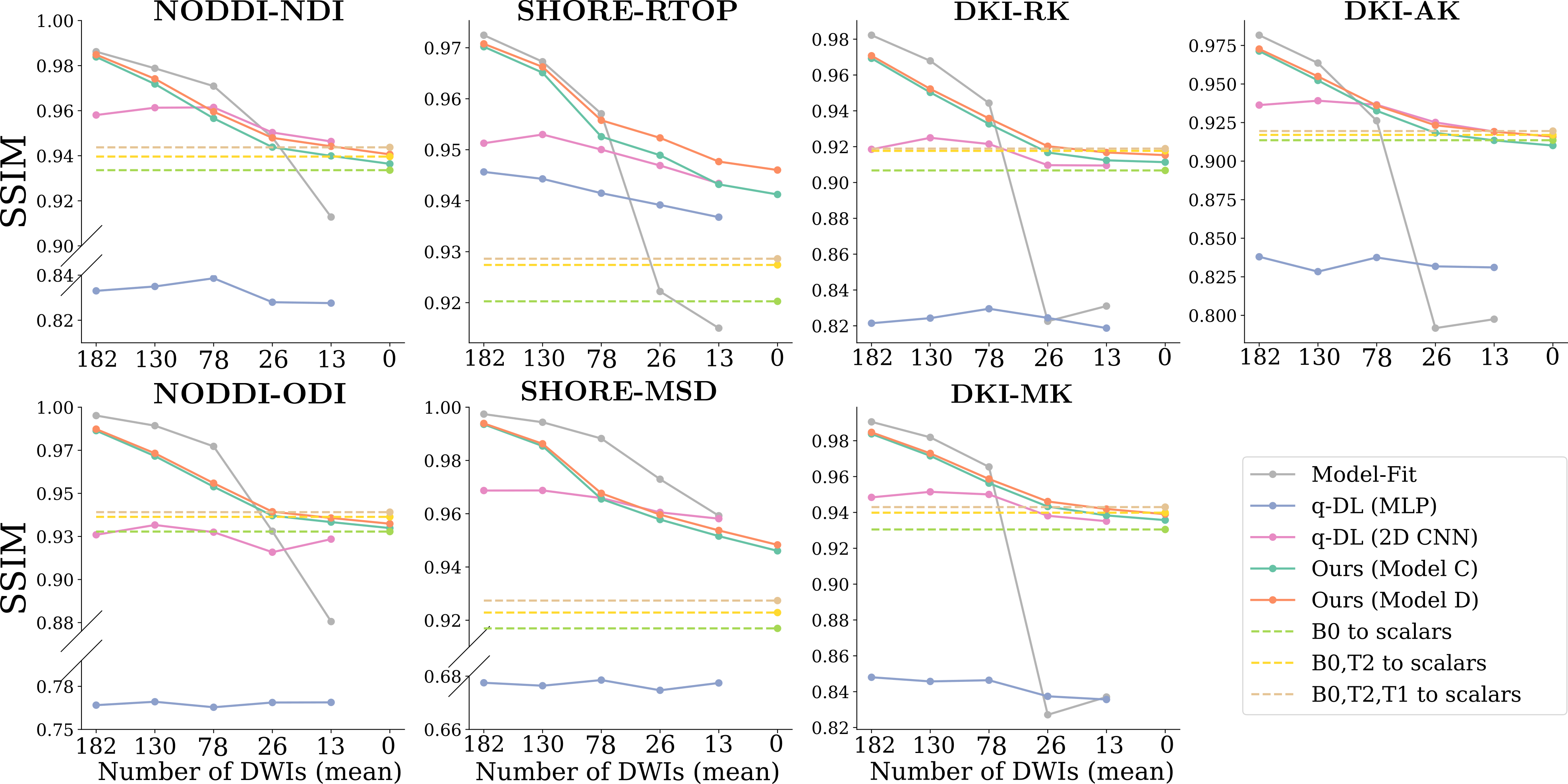}
    \caption{Quantitative microstructural estimation quality under varying DWI acquisition budgets as measured by SSIM (higher is better). }
    \label{fig:ssim-scalar}
\end{figure}

\begin{figure}[t]
    \centering
    \includegraphics[width=1\textwidth]{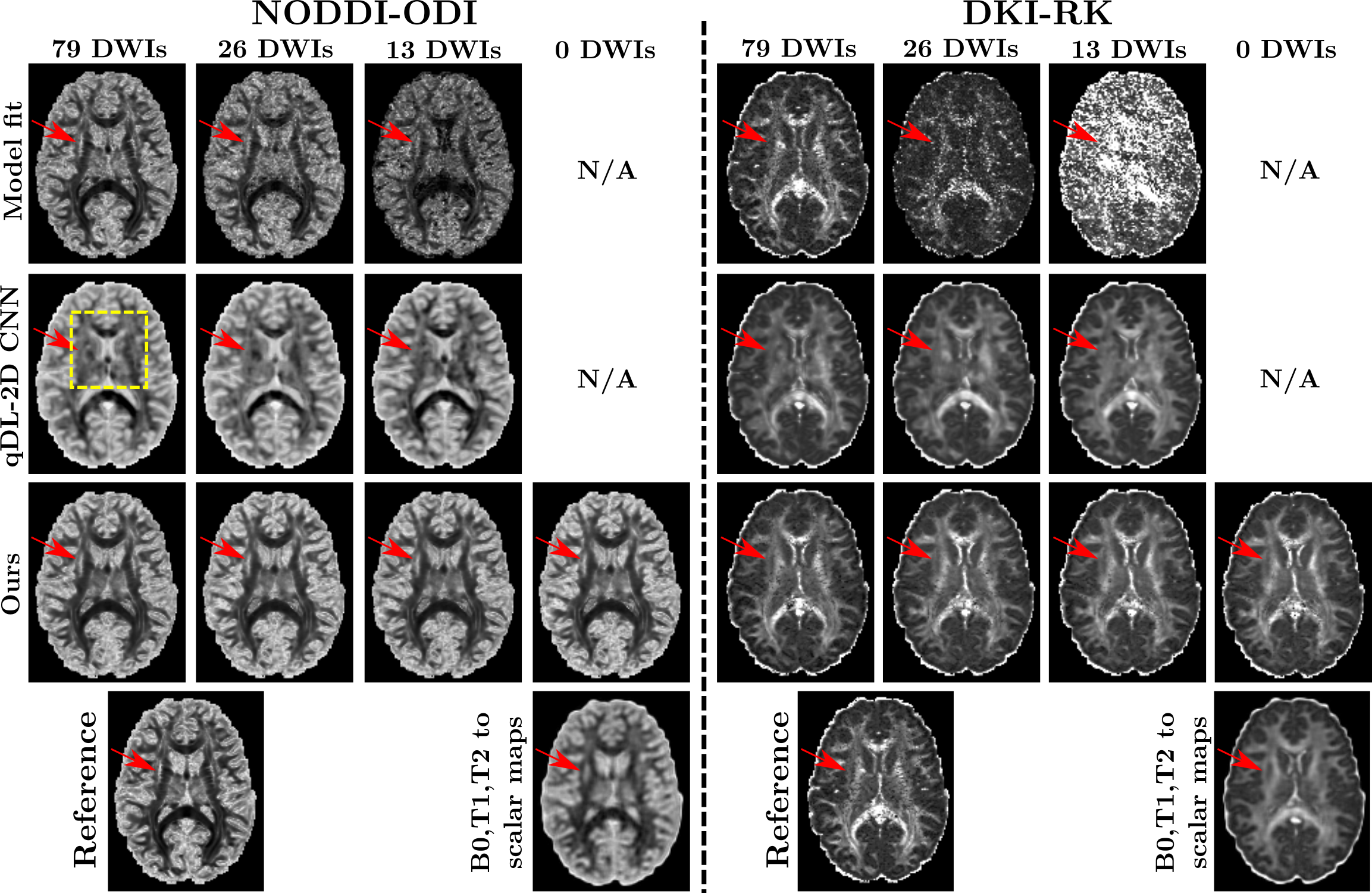}
    \caption{NODDI (ODI) and DKI (RK) scalar maps estimated by different methods from arbitrarily downsampled DWIs. As the number of DWIs decreases, model-fitting results show strong quality degradation and qDL-2D produces inaccurate structural details (see yellow insets), whereas our method preserves microstructure with high fidelity (see external capsule/red arrows). Under fully synthetic settings (0 DWIs), our result displays improved fidelity compared to direct B0,T2,T1 to scalar translation.}
    \label{fig:scalarmap_vis}
\end{figure}

\noindent\textbf{Simulated dense $q$-space for downstream diffusion model fitting.}
Here, we evaluate simulated dense $q$-space sampling by merging synthesized DWIs with the original sparsely-sampled DWIs for the downstream estimation of microstructural indices, tractograms, and dODFs. We randomly downsample in $q$-space at rate $r$ and evaluate the accuracy of downstream estimations.
On held-out test subjects, we downsample in $q$-space such that $k=(1-r)*N$ DWIs are retained. As each subject has different numbers of gradient volumes, we assume that the gradient tables are independent across subjects and thus downsample $q$-space subject-wise with $r=30\%, 50\%, 70\%, 90\%,$ and $95\%$.

\begin{figure}[t]
    \centering
    \includegraphics[width=1.\textwidth]{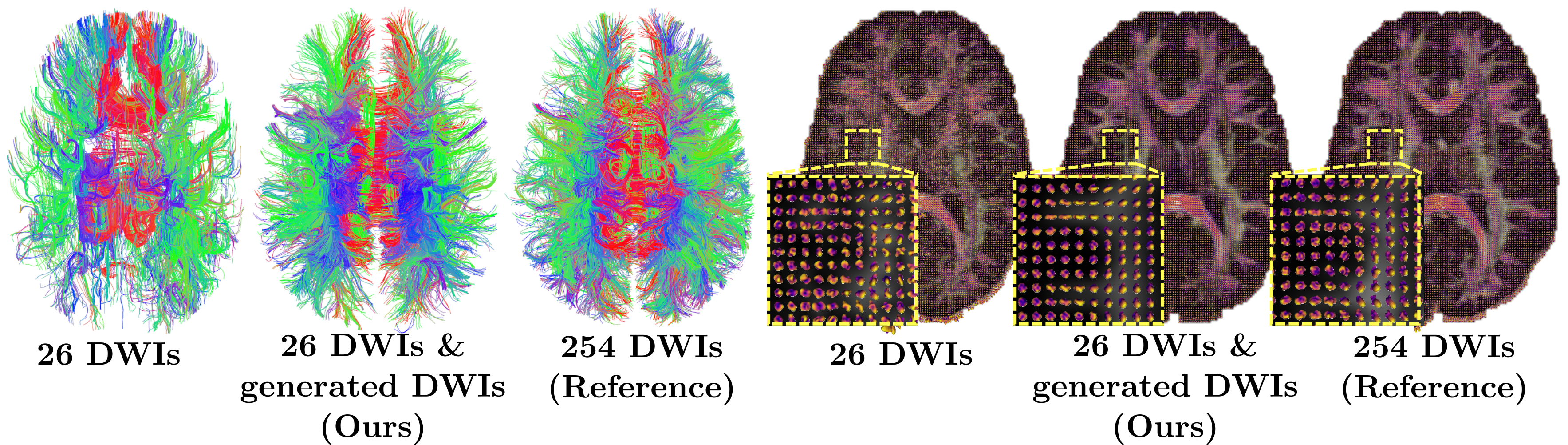}
    \caption{Whole brain fiber-tractography and estimated dODFs overlaid on generalized fractional anisotropy maps calculated from 26 sparsely-sampled DWIs (left), the results from our framework (center), and results from fully-sampled DWIs (right).}
    \label{fig:tract_dodf}
\end{figure}

We benchmark our methods against 3 model-free learning frameworks for the estimation of seven different microstructural indices: NDI, ODI (NODDI~\cite{ZHANG20121000,harms2017robust}); AK, MK, RK (DKI~\cite{jensen2005diffusional,veraart2013weighted}); and RTOP, MSD (SHORE~\cite{ozarslan2008simple,garyfallidis2014dipy}). The derived diffusion maps from both the original DWIs and the synthesized DWIs are calculated and compared with the scalar maps calculated from full DWIs as reference. Compared methods include: (1) q-DL (MLP)~\cite{qspacedl}, (2) q-DL (2D CNN)~\cite{gibbons2019simultaneous}, and (3) a network trained to regress scalar maps from structural MRI inputs following~\cite{gu2019generating,schilling2019synthesized,son2019synthesizing}. (1) and (2) learn scalar maps from the downsampled DWIs, while (3) is implemented as a benchmark to verify our claim that synthesizing DWIs prior to estimating microstructural maps is beneficial both in terms of flexibility and fidelity. To implement (3), we remove all conditioning mechanisms in our generator and experiment with three input settings (B0 only; B0 and T2; B0, T2, and T1) for generating the desired 7-channel scalar maps.

Fig.~\ref{fig:ssim-scalar} presents SSIM between scalar maps estimated by the compared methods and the fully-sampled ground truth. As expected, under lower downsampling rates, standard diffusion model-fits show strong similarity to their fully-sampled references as they are calculated by the same methods.
However, with increased downsampling, model-fitting performance drops dramatically, as also qualitatively shown in Fig.~\ref{fig:scalarmap_vis}. q-DL (MLP)~\cite{qspacedl} does not yield feasible results for this use-case due to its reliance on a predefined downsampling scheme and as MLPs cannot fit diffusion signals which change their input ordering subject-to-subject. Significantly improved performance was obtained with q-DL (CNN)~\cite{gibbons2019simultaneous}. However, q-DL (CNN) still inaccurately predicts structural details (see yellow box in Fig.~\ref{fig:scalarmap_vis}). The three dashed lines in Fig.~\ref{fig:ssim-scalar} present the direct structural MRI to scalar map translation results, with performance gradually increasing with more input modalities. Qualitatively, the produced images lose high-frequency detail as compared to the reference, whereas ours does not. Further, our method possesses increased generalizability across arbitrary downsampling schemes and enables enhanced scalar map estimation with standard model fits.

To summarize, at low downsampling rates, our method achieves competitive results with model-fitting and as the downsampling rate increases, our method generally outperforms other methods. When fully synthetic DWIs are used for generating scalar maps (the 0 DWI setting), our method achieves competitive or better SSIM as compared to the scalar translation framework, alongside signficantly improved qualitative reconstruction as shown in Fig.~\ref{fig:scalarmap_vis}. We note that while we use SSIM to quantify image similarity, it may not correlate exactly with human perception~\cite{nilsson2020understanding}.

To demonstrate further downstream utility beyond microstructural indices, we calculate tensor-based fiber tractography~\cite{jiang2006dtistudio} and SHORE-based diffusion orientation distribution functions (dODF)~\cite{ozarslan2008simple} for a randomly selected subject. Fig.~\ref{fig:tract_dodf} presents qualitative results from downsampled DWIs and the $q$-space restored DWIs in comparison to the fully-sampled reference. While the downsampled results are noisy and inaccurate, the results from our method yield higher fidelity and coherence with respect to the reference.

Further supplemental results and animations including continuous gradient direction interpolations between synthesized DWI are available at \url{https://heejongkim.com/dwi-synthesis}.

\section{Conclusion}
We present a novel $q$-space conditioned translation framework for synthesizing high-fidelity DWIs at arbitrary $q$-space coordinates from multimodal structural MRIs. The model outputs can be merged with downsampled DWIs to simulate dense $q$-space sampling, enabling the usage of advanced diffusion models which require high angular resolution. Some open questions exist, as we train and evaluate on a healthy population under relatively controlled imaging settings. As different imaging modalities may reveal disparate anatomical structure and diffusion properties (e.g., via lesions), the proposed framework requires further validation on larger and more neuroanatomically diverse cohorts. Additionally, real-world considerations such as inter-modality misregistration, resolution differences, and skull stripping discrepancies in multi-modal structural inputs may further impede translation performance. Fortunately, reasonable performance is achieved even when only a single structural modality is available.

\subsubsection*{Acknowledgements}
Work supported by NIH grants 1R01DA038215-01A1, R01-HD055741-12, 1R01HD088125-01A1, 1R01MH118362-01, 1R34DA050287, R01MH122447, and R01ES032294.

\subsubsection*{Conflict of Interest Statement} The authors declare that there are no conflicts or commercial interest related to this article. 

\bibliographystyle{splncs04}
\bibliography{reference}

\newpage
\section*{Supplementary Material}

\begin{table*}[h]
\centering
\begin{tabular}[t]{l|c|c}
\hline
Software & Microstructure maps   & Version    \\ \hline
Diffusion Imaging in Python (Dipy) \cite{garyfallidis2014dipy} & MSD, RTOP  & 1.3.0   \\ \hline
 Microstructure Diffusion Toolbox (MDT) \cite{harms2017robust}  & NDI, ODI &   1.2.6  \\\hline
 \begin{tabular}[c]{@{}l@{}}Diffusion parameter EStImation with \\ Gibbs and NoisE Removal (DESIGNER)\end{tabular} \cite{veraart2013weighted}  & AK, MK, RK &  N/A \\ \hline
\end{tabular}
\caption{Software used to calculate microstructural maps.}
\end{table*}

\begin{figure}[hb]
    \centering
    \includegraphics[width=\textwidth]{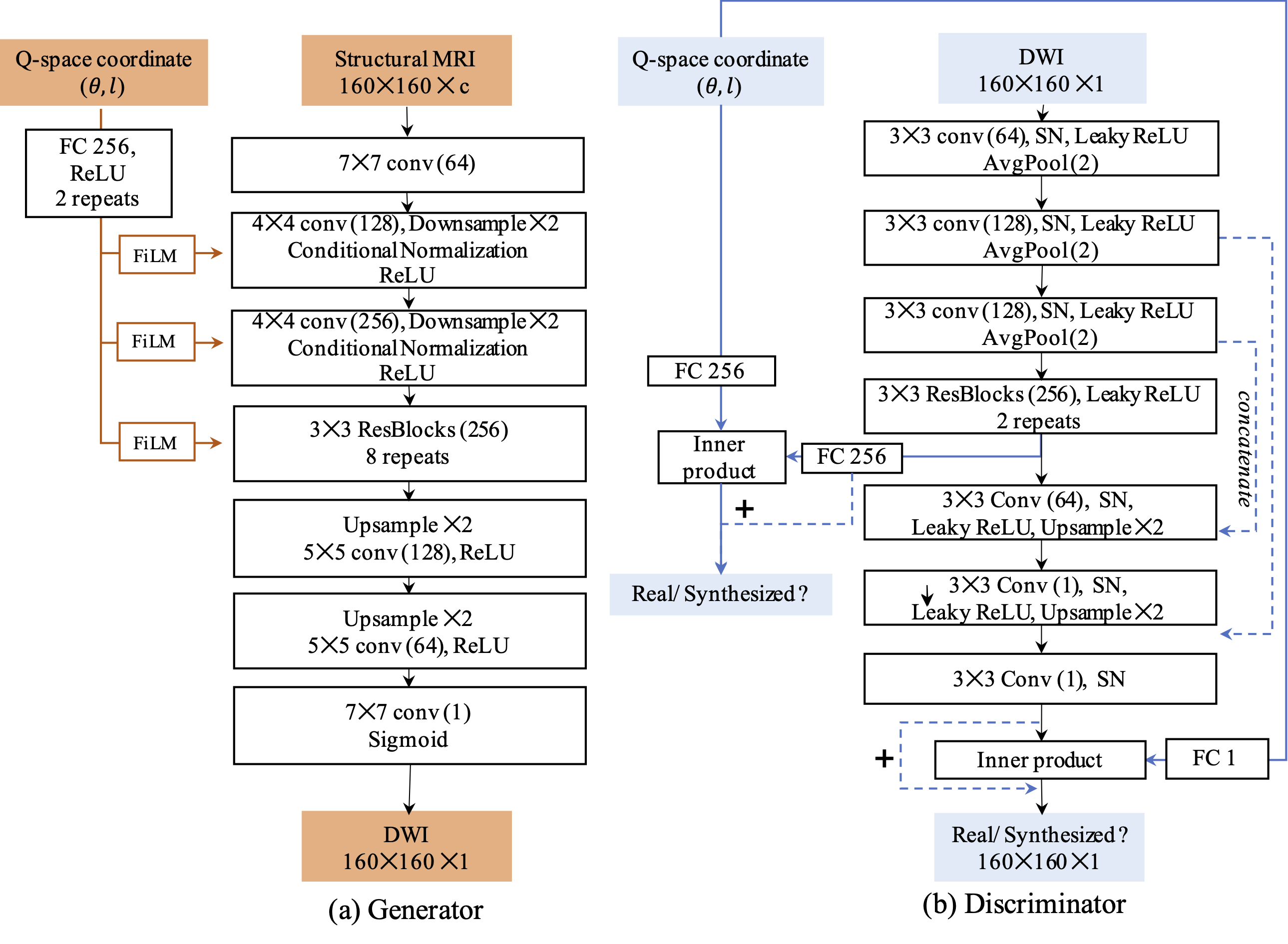}
    \caption{Architecture of the generator and discriminator. SN: Spectral Normalization. }
    \label{fig:arch}
\end{figure}

\begin{figure}
\includegraphics[width=\textwidth]{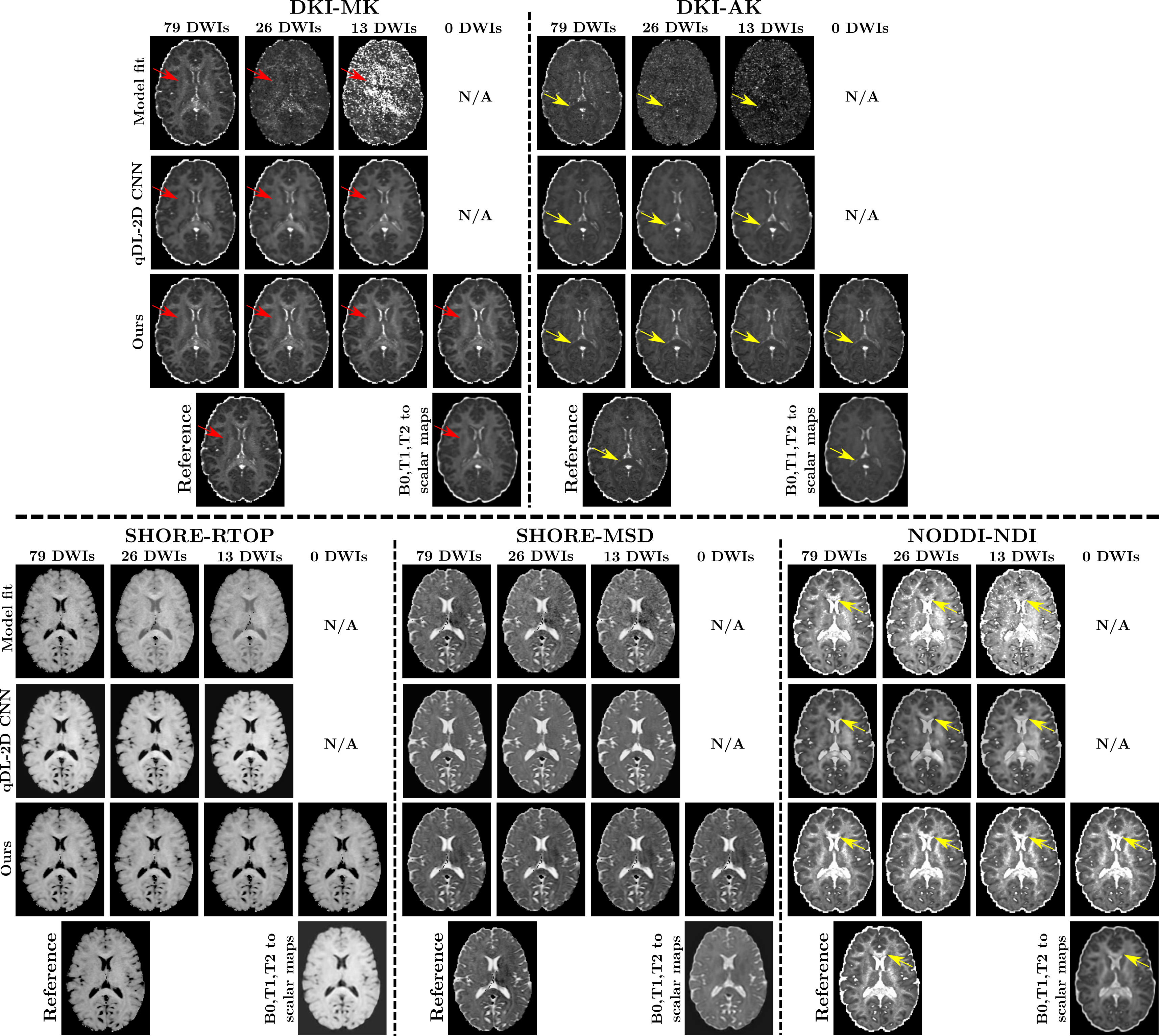}%
\caption{Microstructural maps from diffusion kurtosis imaging (DKI), simple harmonic oscillator based reconstruction and estimation (SHORE) and neurite orientation dispersion and density imaging (NODDI). Our results shows comparable results with the reference in contrast and structures. Model fitting results degrade as the number of available DWIs decreases. qDL-2D prediction predicted insufficient structures, especially external capsule (red arrows) and corpus callosum (yellow arrows). Results from B0,T1,T2 to scalar maps shows overall blurry results while our method well-preserved the structures. Readers are encouraged to zoom in for details.}
\end{figure}

%

\end{document}